\documentclass[aps,preprint]{revtex4}%
\usepackage{amsfonts}
\usepackage{amsmath}
\usepackage{amssymb}
\usepackage{graphicx}%
\begin{document}
\preprint{ }
\title[ ]{On the Observability of Pseudoscalar Toponium at Future Hadron Colliders}
\author{O. Cak\i r$^{1}$, R. Ciftci$^{2}$, E. Recepoglu$^{1}$, S. Sultansoy$^{2,3}$}
\affiliation{$^{1}$Physics Department, Faculty of Sciences, Ankara University, 06100
Tandogan, Ankara, Turkey, $^{2}$Physics Department, Faculty of Sciences and
Arts, Gazi University, 06500 Teknikokullar, Ankara,Turkey, $^{3}$Institute of
Physics, Academy of Sciences, H. Cavid Avenue 33, Baku, Azerbaijan}
\keywords{quarkonia, isosinglet quarks, standard model}
\pacs{}

\begin{abstract}
Possible existence of extra SM families and/or isosinglet $E_{6}$ quarks may
sufficiently decrease $\left\vert V_{tb}\right\vert $ of CKM matrix so that
toponia can be formed. Production of pseudoscalar toponium at the LHC and VLHC
has been considered. The observability of $\ \eta_{t}\rightarrow\gamma\gamma$
signal is discussed.

\end{abstract}
\volumeyear{year}
\volumenumber{number}
\issuenumber{number}
\eid{ }
\date[Date text]{date}
\received[Received text]{date}

\revised[Revised text]{date}

\accepted[Accepted text]{date}

\published[Published text]{date}

\startpage{1}
\endpage{ }
\maketitle

Large value of the top quark mass and estimation of $\left\vert V_{tb}%
\right\vert $ close to one in the case of three SM families have caused
toponium to drop from agenda of particle physics. However, even though mass of
the top quark is fixed, $\left\vert V_{tb}\right\vert $ may be relaxed in such
a way that toponium can be formed. For example, this can be case if extra SM
families exist or $E_{6}$ model is prefered by nature. According to
\cite{Gilman} in the case of three SM generations allowed region for $V_{tb}$
is $0.9990<\left\vert V_{tb}\right\vert <0.9993$, whereas this range becomes
$0.08<\left\vert V_{tb}\right\vert <0.9993$ if there are more than three generations.

It is well-known, that the fundamental fermions masses and mixings and even
the number of fermion generations are not fixed by the Standard Model (SM). In
this sense, SM may be deliberated as an effective theory of fundamental
interactions rather than fundamental particles. The statement of the Flavor
Democracy (or, in other words, the Democratic Mass Matrix approach), which is
quite natural in the SM framework, may be considered as the interesting step
in the true direction \cite{Harari,Fritzsch87,Fritzsch90,Fritzsch94}. It is
intriguing that, Flavor Democracy favors the existence of the fourth standard
model family \cite{Datta,Celikel,Atag,Sultansoy}. The experimental lower
limits on masses of the fourth family fermions are \cite{Hagiwara}:
$m_{\nu_{4}}>45$ GeV from LEP 1, $m_{l_{4}}>100$ GeV from LEP 2, $m_{d_{4}%
}>199$ GeV (neutral current decays, $d_{4}\rightarrow qZ$) and $m_{d_{4}}>128$
GeV (charged current decays, $d_{4}\rightarrow qW$) from FNAL (Tevatron Run
I). Recently \cite{He,Okun} it was shown that a single extra chiral family
with a constrained spectrum is consistent with latest presicion data without
requiring any other new physics source. Moreover, two and three extra
generations with relatively "light" neutrinos ($m_{N}\thickapprox50$ $GeV$)
are also allowed \cite{Okun}.

Another way to relax the condition $\left\vert V_{tb}\right\vert
\thickapprox1$ is the introduction of exotic fermions. We consider as an
example the extension of the SM fermion sector which is inspired by $E_{6}$
GUT model initially suggested by F. Gursey and collaborators
\cite{Gursey,GurseyS}. It is known that this model is strongly favored in the
framework of SUGRA (see \cite{Hewett} and references therein).

\bigskip In hadron collisions, gluon-gluon fusion is the main process for the
production of heavy quarkonia \cite{Barger87}: the $J^{PC}=0^{-\text{ }+}$
pseudoscalar quarkonium state $\eta_{t}(^{1}S_{0})$, which is produced in the
subprocess $gg\rightarrow\eta$, has a production cross section two orders of
magnitude larger than the $J^{PC}=1^{-\text{ }-}$ vector state $\Psi_{t}$,
since $gg\rightarrow g\Psi$ will be the mechanism for the vector quarkonium.
For this reason, lepton colliders with $\sqrt{s}\simeq350$ $GeV$ will be more
suitable for investigation of vector $\psi_{t}$ quarkonium, whereas hadron
machines are best for the investigation of pseudoscalar $\eta_{t}$ quarkonium.

In this work, we consider the process $pp\rightarrow\eta_{t}X$ for the
production of pseudoscalar $(t\overline{t})$ quarkonium with subsequent
$\eta_{t}\rightarrow\gamma\gamma$ decay at the LHC. Unfortunately, the
observability of the $\eta_{t}$ at the Tevatron is out of the question even
for $L^{int}=10$ $fb^{-1}$.

The cross section for $\eta_{t}$ production at hadron colliders can be
expressed as%

\begin{equation}
\sigma\left(  pp\rightarrow\eta_{t}X\right)  =K\frac{\pi^{2}}{8m_{\eta_{t}%
}^{3}}\Gamma\left(  \eta_{t}\rightarrow gg\right)  \tau\int_{\tau}^{1}%
\frac{dx}{x}g\left(  x,Q^{2}\right)  g\left(  \frac{\tau}{x},Q^{2}\right)  ,
\end{equation}
where \cite{Fabiano94}%

\begin{equation}
\Gamma(\eta_{t}\rightarrow gg)=\frac{8\alpha_{s}^{2}\left(  Q^{2}\right)
}{3m_{\eta_{t}}^{2}}\left\vert R_{s}\left(  0\right)  \right\vert ^{2}\left[
1+\frac{\alpha_{s}\left(  Q^{2}\right)  }{\pi}\left(  \frac{\pi^{2}}{3}%
-\frac{20}{3}\right)  \right]  ,
\end{equation}
$\alpha_{s}\left(  Q^{2}\right)  $ is the strong coupling constant and
$\tau=m_{\eta_{t}}^{2}/s$ with $\sqrt{s}$ being the center of mass energy of
the collider. $R_{s}\left(  0\right)  $ is the radial wave function of the
S-state evaluated at the origin \cite{Barger87}. $K\thickapprox1.4$ is the
enhancement factor for the next-to-leading order QCD effects \cite{Kuhn}. For
gluon distrubition function $g\left(  x,Q^{2}\right)  $ we have used CTEQ5L
\cite{CTEQ} with $Q^{2}=m_{t}^{2}$.

\bigskip The main decay modes of $\eta_{t}$ are the single quark decay and
$\eta_{t}\rightarrow gg$. The branching ratio for $\eta_{t}\rightarrow
\gamma\gamma$ is given by%

\begin{equation}
Br\left(  \eta_{t}\rightarrow\gamma\gamma\right)  \approx\frac{\Gamma\left(
\eta_{t}\rightarrow\gamma\gamma\right)  }{\Gamma\left(  \eta_{t}\rightarrow
gg\right)  +2\Gamma_{t}},
\end{equation}
where [\cite{Mirkes92}, \cite{Fabiano98}]%

\begin{equation}
\Gamma\left(  \eta_{t}\rightarrow\gamma\gamma\right)  =\frac{12Q_{t}^{4}%
\alpha^{2}\left(  Q^{2}\right)  }{m_{\eta_{t}}^{2}}\left\vert R_{s}\left(
0\right)  \right\vert ^{2}\left[  1+\frac{\alpha_{s}\left(  Q^{2}\right)
}{\pi}\left(  \frac{\pi^{2}}{3}-\frac{20}{3}\right)  \right]  ,
\end{equation}

\begin{equation}
\Gamma_{t}=\frac{m_{t}^{3}}{16\pi v^{2}}\left\vert V_{tb}\right\vert
^{2}\left[  1-\left(  \frac{m_{W}}{m_{t}}\right)  ^{2}\right]  ^{2}\left[
1+2\left(  \frac{m_{W}}{m_{t}}\right)  ^{2}\right]  \left[  1-\frac{2}{3}%
\frac{\alpha_{s}\left(  Q^{2}\right)  }{\pi}\left(  \frac{2\pi^{2}}{3}%
-\frac{5}{2}\right)  \right]  ,
\end{equation}
where $\alpha\left(  Q^{2}\right)  $ is fine-structure constant, $v\approx245$
$GeV$ is the vacuum expectation value of the Higgs field.

For the background calculations we use COMPHEP 4.2 \cite{Comphep}%
.\bigskip\ The dominant backgrounds are $f\overline{f}\rightarrow\gamma\gamma$
and $gg\rightarrow\gamma\gamma$ with cross sections $2\times10^{4}$ pb and
$3\times10^{5}$ pb, respectively. In order to suppress the backgrounds we
apply a cut $p_{T}>0.4m_{\eta_{t}}$ on transverse momentum of both photons.
This requirement reduces the signal by $\sim40\%$, whereas the background
drops drastically. Furthermore, we use $60\%$ efficiency for two photon
identification. Finally, we consider a mass window $m_{\gamma\gamma}\pm
2\sigma_{m}$ for two photons invariant mass using%

\begin{equation}
\sigma_{m}=m_{\gamma\gamma}\left(  \frac{0.07}{\sqrt{E_{\gamma}}%
}+0.005\right)  .
\end{equation}
As a result we obtain the signal cross sections for $\left\vert V_{tb}%
\right\vert =0.1$, $0.2$ and $0.3$ shown in the fifth column of the Table I.
The signal cross-sections for $\eta_{t}\rightarrow\gamma\gamma$\ channel are
practically insensitive to the Higgs mass in the range $120<m_{H}<250$
GeV.\ Using the criteria given above we calculate the background cross
sections presented in the third column of the table.

We have evaluated the statistical significance for the signal using%

\begin{equation}
S=\frac{N_{S}}{\sqrt{N_{S}+N_{B}}}%
\end{equation}
The statistical significance for $L^{int}=1000$ $fb^{-1}$(characteristic value
for superbunch options \cite{Takayama} of LHC and VLHC) and the integrated
luminosities needed to achieve $3\sigma$ and $5\sigma$ discovery criteria are
presented in the last three columns of the Table I. 

In conclusion, pseudoscalar toponium could manifest itself at the LHC with
$\sqrt{s}=14$ TeV if $\left\vert V_{tb}\right\vert \approx0.1$. A possible
upgrade of the LHC with $\sqrt{s}=28$ TeV \cite{Azuclos} will give opportunity
to reach $\left\vert V_{tb}\right\vert \approx0.2$.\ If $\left\vert
V_{tb}\right\vert \lesssim0.25$, $\eta_{t}$ could be observed at VLHC with
$\sqrt{s}=40$ TeV, whereas VLHC with $\sqrt{s}=175$ TeV cover whole range up
to $\left\vert V_{tb}\right\vert \approx0.4$. Finally, let us mentioned, that
measurement of $V_{tb}$ via s-channel single top at ATLAS \cite{Pineiro} will
as well give an indirect information on the existence of the pseudoscalar
toponium. \ \bigskip

Authors are grateful to A. K. Ciftci and S. A. \c{C}etin for useful
discussions. This work is supported in part by Turkish Planning Organization
(DPT) under the Grant No 2002K120250.

\bigskip

\begin{table}[b]
\caption{$\eta_{t}\rightarrow\gamma\gamma$ channel: The number of signal,
background events and corresponding statistical significancens for
$L_{int}=1000$ fb$^{-1}$. The integrated luminosities needed to achieve
$3\sigma$ and $5\sigma$ levels are also given.}
\begin{center}%
\begin{tabular}
[c]{|c|c|c|c|c|c|c|c|}\hline
& $\sqrt{s}$ (TeV) & $\sigma_{B}$ (fb) & $V_{tb}$ & $\sigma_{S}$ (fb) &
$S/\sqrt{B}$ & $L_{int}$(fb$^{-1}$) for $3\sigma$ & $L_{int}$(fb$^{-1}$) for
$5\sigma$\\\hline
&  &  & 0.1 & 0.30 & 2.36 & 1600 & 4500\\
& 14 & 16.14 & 0.2 & 0.08 & 0.63 & 22700 & 63000\\
LHC &  &  & 0.3 & 0.04 & 0.31 & 90800 & 252000\\\cline{2-8}
&  &  & 0.1 & 1.29 & 8.72 & 120 & 330\\
& 28 & 21.90 & 0.2 & 0.35 & 2.36 & 1600 & 4500\\
&  &  & 0.3 & 0.16 & 1.08 & 7700 & 21400\\\hline
&  &  & 0.1 & 2.54 & 14.13 & 45 & 125\\
& 40 & 32.30 & 0.2 & 0.70 & 3.89 & 600 & 1600\\
VLHC &  &  & 0.3 & 0.31 & 1.72 & 3000 & 8400\\\cline{2-8}
&  &  & 0.1 & 22.71 & 60.37 & 2.5 & 7\\
& 175 & 141.50 & 0.2 & 6.28 & 16.69 & 32 & 90\\
&  &  & 0.3 & 2.82 & 7.50 & 160 & 450\\\hline
\end{tabular}
\end{center}
\end{table}

\end{document}